# Giant-Magnetocaloric effect and phonon dynamics in (GdCe)CrO$_3$

Ravi Kiran Dokala[1,3†*] Shaona Das[2,3†*], and Subhash Thota[3*]

[1]*Solid State Division, Department of Materials Science and Engineering, Uppsala University, Uppsala-75237, Sweden*
[2]*Institute of Material Science, Technical University of Darmstadt, Darmstadt-64287, Germany*
[3]*Department of Physics, Indian Institute of Technology Guwahati, Guwahati, Assam-781039, India*

**Abstract**

We investigate the effect of Ce$^{3+}$ substitution on the magnetic ordering and phonon dynamics of the GdCrO$_3$ orthorhombic perovskite. The Ce doped compound exhibits long-range canted antiferromagnetism with Néel transitions, $T_N$ at ~173 K, accompanied by spin-flip, $T_{SF}$ at ~ 10 K. Ce$^{3+}$ incorporation drives a modification of the spin-flip transition from the $\Gamma_4(G_x, A_y, F_z)$ configuration to $\Gamma_4'$ inducing a reorientation of the spin axis between the $(00\bar{1})$ and $(001)$ crystallographic planes. This spin reorientation is governed by Zeeman energy and produces pronounced field-induced irreversibility between FCC and FCW magnetization processes. The substituted compound Gd$_{0.9}$Ce$_{0.1}$CrO$_3$ (GCCO) exhibits a remarkably large magnetic entropy change, $\Delta S_M$ ~ 45-40 J/kg-K for $\Delta H$ = 90-70 kOe at 3 K among the highest reported for rare-earth orthochromites. The interplay of spin-only magnetocrystalline anisotropy from Cr$^{3+}$ and spin–orbit–driven magnetic moments of Gd$^{3+}$ and Ce$^{3+}$ results in pronounced spin–phonon coupling, manifested through the A$_{1g}$(6) vibrational mode. The observed temperature-dependent spectral evolution is consistent with behaviour reported in isostructural magnetic perovskites.

[†]equal contribution

[*]Corresponding authors: Ravi Kiran Dokala, ravi.kiran.dokala@angstrom.uu.se
Shaona Das, shaona.das@oxide.tu-darmstadt.de
Subhash Thota, subhasht@iitg.ac.in

1. **Introduction**

Heavier rare-earth ($R$) perovskites, particularly the $R$-chromates $R$CrO$_3$ (R = Pr, Ce, Gd, Sm, etc.), exhibit highly complex magnetic ordering characterized by robust spin dynamics and strong exchange interactions among Cr$^{3+}$–Cr$^{3+}$, $R^{3+}$–Cr$^{3+}$ and $R^{3+}$–$R^{3+}$ sublattices [1, 2]. Owing to these interactions, the chromate family demonstrates several intriguing physical phenomena such as spin-flip transitions, negative magnetization and compensation effects, multiferroicity, giant magnetocaloric response, and field-induced magnetic-phase transitions, which are of significant interest for the design and development of spin-valve and related spintronic devices [1, 2]. Rare-earth orthochromites have been extensively investigated over the past several decades using a wide range of experimental techniques including heat capacity measurements, neutron diffraction, Mössbauer spectroscopy, magnetodielectric studies, and optical-absorption spectroscopy to probe their emergent ferroic and magnetic states [3–5]. These studies have demonstrated that the coexistence of local non-centrosymmetric structural distortions and multiple magnetic-ordering components (antiferromagnetic and ferromagnetic) plays a crucial role in governing the ferroelectric response in RCrO$_3$. In addition, phenomena such as exchange bias, spin-reorientation transitions, negative magnetization and its reversal, and spin-glass-like behavior have also been widely reported in $R$CrO$_3$-based systems [4–8]. Among these effects, magnetization reversal is particularly captivating due to its origin in the contrasting temperature dependencies of magnetic moments associated with distinct magnetic sublattices. In such compounds, the total magnetization switches sign from positive to negative below a characteristic temperature known as the compensation temperature ($T_{\text{Comp}}$), a feature now regarded as a hallmark of $R$CrO$_3$ systems [9]. Magnetization reversal, which relies on the presence of two competing magnetization components, renders these materials promising for thermally or magnetically driven switching technologies, including nonvolatile magnetic memory storage, thermo-magnetic switches, high-speed read/write magnetic memory, and thermally assisted MRAM architectures [10].

Most $R$CrO$_3$ compounds, however, exhibit $T_{\text{Comp}}$ significantly below room temperature, which limits their applicability in practical magneto-electronic devices. Consequently, enabling the tunability of $T_{\text{Comp}}$ toward higher temperatures has become an important research objective. The internal molecular field generated by the transition-metal sublattice (Cr$^{3+}$, Mn$^{3+}$, etc.) plays a decisive role in driving magnetization reversal in $R$CrO$_3$-type materials: rare-earth ions (e.g. Gd$^{3+}$, Nd$^{3+}$, Sm$^{3+}$, and Ce$^{3+}$) couple antiferromagnetically with the transition-metal ions, resulting in polarization and alignment of their magnetic moments opposite to the external field, ultimately yielding a net negative magnetization [11]. For most members of the $R$CrO$_3$ series (e.g. PrCrO$_3$, YbCrO$_3$, HoCrO$_3$), magnetization reversal is observed in the field-cooled warming (FCW) protocol. Remarkably, the end-member compound GdCrO$_3$ displays negative magnetization only during the field-cooled cooling (FCC) cycle, highlighting the strong dependence of the magnetic response on the measurement history and protocol [12].

In addition to their diverse magnetic characteristics, complex oxide chromates are widely recognized for exhibiting a substantial magnetocaloric effect (MCE), making them promising candidates for environmentally benign magnetic refrigeration technologies. Such systems offer an attractive alternative to conventional gas-compression refrigeration, which continues to dominate global markets despite its environmental hazards [13]. The magnetic interactions driving exotic phase transitions within the constituent magnetic sublattices enable these materials to operate as refrigerants across a broad temperature range. Depending on the transition temperatures involved, chromates hold strong potential for hydrogen and helium liquefaction, thereby competing effectively with well-established intermetallic MCE systems such as $GdNi_2$ and $La(Fe,Si)_{13}$ [14, 15]. Previous studies have reported a giant MCE in polycrystalline $GdCrO_3$ with $-\Delta S_M \sim 36.97$ J.Kg$^{-1}$K$^{-1}$, the magnitude of $-\Delta S_M$ can be further enhanced by tailoring selective magnetic interactions that reinforce the overall magnetic moment [4].

Among chromates, $GdCrO_3$ and $CeCrO_3$ are prototypical antiferromagnetic systems well known for displaying magnetization-reversal behavior [3, 4]. $GdCrO_3$ is a G-type AFM compound with $T_N \sim 169$ K and a compensation temperature $T_{Comp} \sim 132$ K; a spin-flip transition occurs at $T_{SF} \sim 18$ K at fields as low as 100 Oe. A giant MCE has also been reported at low temperatures with $-\Delta S_M \sim 36.97$ J.Kg$^{-1}$K$^{-1}$ at 70 kOe. $CeCrO_3$ also exhibits a G-type AFM spin configuration, though its spin-flip transition takes place at higher magnetic fields and ultimately compensates the magnetization, driving it into the positive regime. According to the report by Cao et al., $CeCrO_3$ displays $T_N \sim 230$ K, $T_{Comp} \sim 100$ K, and $T_{SF} \sim 36$ K at 1.2 kOe [3].

In the present work, we investigate the polycrystalline chromate perovskite system GCCO by examining its structural, electronic, and magnetic properties. X-ray diffraction and Raman spectroscopy provide insights into the internal atomic structure and its response to vibrational excitations. The GCCO system demonstrates magnetization reversal, a field-tunable spin-flip transition, and exchange bias—indicative of multiple coexisting magnetic configurations. Furthermore, GCCO exhibits a giant magnetocaloric effect, attributed primarily to local structural distortions introduced by substituting trivalent $Ce^{3+}$ at the relatively smaller $Gd^{3+}$ sites.

2. **Methods**

Polycrystalline GCCO was synthesized by a standard solid-state reaction route. Stoichiometric amounts of $Gd_2O_3$ (99.9%), $CeO_2$ (99.95%), and $Cr_2O_3$ (99.99%) were thoroughly ground in an agate mortar for 5 h and calcined at 1000 °C for 24 h in air to ensure homogeneity. The calcined powder was reground for 2 h, pelletized, and sintered at 1200 °C for 24 h in a Nabertherm tubular furnace, followed by natural cooling to room temperature. Phase purity was confirmed using room-temperature X-ray diffraction (Rigaku TTRAX III, Cu-K$\alpha_1$, $\lambda = 1.54056$ Å, step size 0.02°). X-ray photoelectron spectroscopy (XPS) was carried out using a PHI 5000 VersaProbe III system with an Al K$\alpha$ source to determine the core-level charge states. DC magnetization measurements were performed using a Quantum Design PPMS (Dyna-Cool model). Magnetization was recorded as a function of temperature (3–300 K) and magnetic field (50 Oe–20 kOe) under zero-field cooled warming (ZFCW), field-cooled cooling (FCC), and field-cooled warming (FCW) protocols. Isothermal M–H loops were measured between ±90 kOe at selected temperatures, and time-dependent magnetization was recorded under the FCC protocol. Room-temperature Raman spectra (100–800 cm$^{-1}$) and low-wavenumber scans (100–200 cm$^{-1}$) were obtained using a Horiba Jobin Yvon LabRam HR spectrometer with a 20 mW He–Ne laser. Temperature-dependent Raman measurements down to 80 K were conducted using a THMS600 module.

3. **Results and Discussion**

GdCrO$_3$ and CeCrO$_3$ typically crystallize in a slightly distorted orthorhombic perovskite structure with space group *Pbnm* [16-19]. In GCCO, the larger Ce$^{3+}$ ions substitute at the *A*-site (12-fold coordination) without altering the *B*-site Cr$^{3+}$ environment (6-fold coordination). Figure 1 shows the XRD patterns of the GCCO samples, confirming single-phase formation without secondary impurities. Rietveld refinement was performed using FULLPROF [20], and all Bragg reflections were indexed to the *Pbnm* (No. 62) structure. The obtained goodness-of-fit values ($\chi^2$ = 2.42 and 2.63) indicate reliable refinement. The *A*-site occupancy of Gd/Ce was refined and found to be consistent with the nominal stoichiometry. Ce substitution leads to slight increases in the lattice parameters *a* and *b*, consistent with the larger ionic radius of Ce$^{3+}$ (1.143 Å) compared to Gd$^{3+}$ (1.053 Å; 8-coordination). Consequently, the unit-cell volume increases by ~0.072%, and the *c* parameter also shows a small positive shift. In order to understand the internal crystallographic behaviour, we tabulated all the structural parameters refined and calculated along with the comparative study of the pristine compounds from the previously reported values in Table 1. These changes can be explained with the help of difference of average radius of *A*-site cation: $r_{avg} = \sqrt{[(0.9) \times R_{Gd^{3+}}^2] + (0.1 \times R_{Ce^{3+}}^2)}$ for GCCO sample which increases to 1.063 Å compared to 1.053 Å for pristine compound with $R_{Gd^{3+}}$ = 1.053 Å having VIII coordination number and that of $R_{Ce^{3+}}$ = 1.143 Å having the same coordination number [21]. The substitution of larger cation impacts the crystal symmetry by introducing significant distortion of the unit-cell which is popularly known as the factor of tolerance (*t*), given in the following Eq. (1).

$$t = \frac{\{[(1-x) \times R_{Gd^{3+}}] + (x \times R_{Ce^{3+}}) + R_{O^{2-}}\}}{\sqrt{(R_{Cr}^{3+} + R_O^{2-})}} \quad 1$$

Accordingly, the octahedral distortion (Δ) can be expressed as below in Eq. (2).

$$\Delta = \left(\frac{1}{N}\right) \Sigma_{n=1,N} \{(d_n - \{d\}) \not< d >\}^2 \quad 2$$

Here, the apical Cr-O$_{(1)}$-Cr bond angle ($\theta_1$), basal Cr-O$_{(2)}$-Cr bond angle ($\theta_2$) and the tilt angles θ and ϕ inside the CrO$_6$ octahedra along the pseudo-cubic axes [110] and [001] calculated from the bond angles that exhibit visible deviations with the *A*-site cation substitution than that of the average bond lengths between the Cr and the O anions (changes ~ 0.54 %) directing towards the bond rigidity expected in octahedral symmetry surrounding the trivalent Cr [22]. Also, we have calculated the tilt angles from the experimentally obtained lattice parameters, $\theta = cos^{-1}\frac{\sqrt{2}a}{b}$, $\varphi = cos^{-1}\frac{a}{b}$ and the magnitudes match well with the previously reported values of similar systems [23, 24]. Along with the tilting of the CrO$_6$ octahedra, due to the mismatch between the A-site Gd/Ce and B-site Cr cations, an impulsive reduction in the strain parameter *s* with an increment of ~ 2.2 % is observed as the average radius of the *A*-site cation comprises the Wyckoff position eccentricities of Gd/Ce (4c) and the Cr (4b) sites.

X-ray photoelectron spectroscopy (XPS) was employed to examine the electronic structure, oxidation states, and surface composition of the synthesized GCCO perovskites. Figure 1 (*i–iv*) presents the room-temperature high-resolution spectra, where the scattered points correspond to experimental data and the solid curves represent Lorentzian–Gaussian fitting results. The spectra were collected with a binding-energy resolution of 0.1 eV, and the background was corrected using the Tougaard method. Peak fitting was performed through a nonlinear least-squares procedure, with the binding-energy scale calibrated to the C 1s peak at 285 eV. Core-level spectra of Gd-4$d$, Ce-3$d$, Cr-2$p$, and O-1$s$ were analyzed individually, and all obtained binding energies matched well with standard NIST values [25]. The Gd-4$d$ spectrum shows two clear spin–orbit doublets corresponding to the 4$d_{5/2}$ and 4$d_{3/2}$ states. Peaks at 139.1, 141.4, 145.3, and 147.4 eV agree with literature values for Gd$^{3+}$ [26–28]. Additional lower-energy components (138.4 and 140.8 eV) also arise from the 4$d_{5/2}$ level. The goodness-of-fit ($\chi^2 \approx 2.7$) confirms reliable peak deconvolution, and the absence of satellite structures indicates that Gd is stabilized in the trivalent state. Figure 1(*ii*) displays the Ce-3$d$ core-level spectrum, containing four major peaks at 882.1, 884.7, 899.3, and 903.5 eV. These features originate from the 3$d_{5/2}$ and 3$d_{3/2}$ components and correspond to Ce$^{3+}$ states [30]. Although Ce frequently exhibits Ce$^{3+}$/Ce$^{4+}$ mixed valence in oxides, no signature of Ce$^{4+}$ was observed in GCCO, confirming the chemical homogeneity and phase purity. Overall, the XPS results validate that all constituent elements retain their expected oxidation states, with no evidence of extrinsic phases or additional oxidized species.

Raman spectroscopy is a sensitive probe of lattice vibrations, octahedral tilts, and local distortions in orthorhombic perovskite oxides, offering insight into cation displacements and bond-strength variations that influence their physical properties. It is particularly effective in detecting subtle symmetry changes arising from cation-size mismatch, such as that between Gd$^{3+}$ (1.053 Å) and Ce$^{3+}$ in GCCO. In rare-earth chromites, Raman-active phonons reflect the stability of the CrO$_6$ octahedra and their coupling to the *A*-site environment. Figure 2(a) shows the temperature-dependent Raman spectra of GCCO (80–293 K, 100–800 cm$^{-1}$). Based on the orthorhombic *Pbnm* structure with Glazer tilt system a$^-$b$^+$a$^-$, group theory predicts 60 modes, of which 24 are Raman active (7A$_{1g}$ + 5B$_{1g}$ + 7B$_{2g}$ + 5B$_{3g}$) [21-23]. These correspond to octahedral stretching, bending, rotational modes, and *A*-site cation motions. As Cr$^{3+}$ (3$d^3$) is Jahn–Teller inactive, distortions mainly arise from octahedral tilts and *A*-site effects [22]. Eleven Raman modes are observed experimentally, showing temperature-dependent red- and blue-shifts indicative of anharmonic phonon interactions.

The low-frequency phonons (< 200 cm$^{-1}$) in orthorhombic chromite perovskites are primarily governed by rare-earth *A*-site cation vibrations, whose frequencies depend on the reduced mass (μ) according to ω = √(k/μ), where k is the effective force constant. In GCCO, these low-energy modes are assigned as A$_{1g}$(1) and B$_{2g}$(1). Above 200 cm$^{-1}$, the phonon modes arise from vibrations involving Gd/Ce and oxygen atoms as well as bending and stretching motions of the Cr–O bonds. Among these, the A$_{1g}$(1) and B$_{2g}$(1) doublets originate from collective octahedral rotations around the y-axis characteristic of the a$^-$b$^+$a$^-$ tilt system. At higher frequencies, the mode at ~ 561 cm$^{-1}$ (80 K) corresponds to CrO$_6$ octahedral bending, while the intense 698 cm$^{-1}$ peak reflects antisymmetric octahedral stretching that is highly sensitive to Cr–O bond-length and bond-angle variations. The stability of these modes, along with their moderate and systematic shifts with temperature, confirms the absence of structural phase transitions across the measured range and supports the robustness of the orthorhombic phase.

Temperature-dependent Raman analysis reveals systematic softening and hardening trends consistent with anharmonic phonon behavior. For GCCO, the $A_{1g}(6)$ mode shows a clear temperature-dependent shift toward lower wavenumbers upon warming, directly linking changes in (Gd/Ce)/Cr–O bond lengths and O–Cr–O bond angles to the vibrational response. Figure 2(b) illustrates the fitted $A_{1g}(6)$ mode near $T_N$, where the peak positions were extracted using Lorentzian fitting. To examine the interplay between lattice vibrations and magnetism, the Raman line-shape parameters were analyzed across the magnetic ordering temperature. A subtle but distinct kink at $T_N$ indicates weak yet measurable spin–phonon coupling (SPC). Similar weak anomalies have been reported in other $R$CrO$_3$ systems, often attributed to the interplay between rare-earth magnetism, Cr–O–Cr superexchange, and structural distortions.

The temperature-dependence of the phonon-mode with frequency ω follows the conventional relation [29]:

$$\omega(T) = \omega_0(T) + \Delta\omega_{lat}(T) + \Delta\omega_{sp-la}(T) + \Delta\omega_{el-ph}(T) + \Delta\omega_{an}(T) \quad \quad 3$$

Here, $\omega_0$ is the harmonic frequency at 0 K, $\Delta\omega_{latt}$ the quasi-harmonic lattice-volume contribution, $\Delta\omega_{sp-lat}$ the spin–lattice coupling term, $\Delta\omega_{el-ph}$ the electron–phonon component (negligible for insulating chromites), and $\Delta\omega_{anh}$ the intrinsic anharmonic contribution. For GCCO, where electron-phonon coupling is insignificant and lattice effects are modest, the frequency evolution reduces to:

$$\omega(T) = \Delta\omega_{sp-lat}(T) + \Delta\omega_{an}(T) \quad \quad 4$$

The dominant mechanism above $T_N$ is the anharmonic phonon-phonon interaction, which leads to frequency hardening as temperature decreases. This contribution is modelled using the standard three-phonon and four-phonon decay processes [24]:

$$\omega_{anh}(T) = \omega_0 + A\left[1 + \frac{2}{e^{\frac{\hbar\omega_0}{2k_BT}} - 1}\right] + B\left[1 + \frac{3}{e^{\frac{\hbar\omega_0}{3k_BT}} - 1} + \frac{3}{(e^{\frac{\hbar\omega_0}{3k_BT}} - 1)^2}\right] \quad \quad 5$$

where, the parameters A and B were obtained by fitting temperatures above $T_N$, and the resulting curves were extrapolated below $T_N$ to highlight deviations clearly attributable to magnetic ordering. As shown in Fig. 2(b), the $A_{1g}(6)$ mode of GCCO exhibits an anomalous softening immediately below $T_N$, followed by hardening at lower temperatures. This behavior reflects the competition between exchange-striction effects (which soften the mode as spins begin to order) and stabilization of the long-range antiferromagnetic state (which increases restoring forces at lower T). The observed ~4 cm$^{-1}$ total shift further confirms this interplay.

To quantify SPC strength, the Granado model is invoked, where phonon renormalization scales with the nearest-neighbour spin correlation [29]:

$$\Delta\omega_{sp-lat} \propto <S_i.S_j> \quad \quad 6$$

In the molecular-field approximation:

$$\Delta\omega_{sp-lat} \approx \lambda S^2 \left[1 - \left(\frac{T}{T_N}\right)^\gamma\right] \qquad 7(a)$$

$$\approx \lambda \left[\frac{M(T)}{M_{sat}}\right]^2 \qquad 7(b)$$

For GCCO, fitting these expressions produced $\lambda$ values fluctuating between 0.3–0.7 cm$^{-1}$, reflecting the non-monotonic phonon behavior arising from the abrupt softening near $T_N$ and subsequent hardening. Such complexity is typical of systems where octahedral tilts and A-site disorder influence local exchange pathways.

Further insight into SPC comes from structural parameters: the CrO$_6$ octahedral distortion $\Delta$, extracted from Rietveld refinement, is $1.28 \times 10^{-4}$ for GCCO, indicating weak static distortion. Even this small $\Delta$ can significantly affect the Cr–O–Cr exchange integral, which is highly sensitive to bond-angle variations (following Goodenough–Kanamori rules). The rare-earth magnetic moments also play a key role: Gd$^{3+}$ ($\mu_{eff}$ = 7.94 $\mu_B$, $L = 0$) contributes primarily spin-only magnetism, whereas Ce$^{3+}$ ($\mu_{eff}$ = 2.54 $\mu_B$, L = 3) introduces moderate spin–orbit coupling. Their combined presence in GCCO results in a weaker net SPC, consistent with the modest Raman anomalies observed. Overall, the Raman analysis confirms that both octahedral distortions and rare-earth magnetism influence spin–phonon interactions in GCCO, with the A$_{1g}$(6) mode serving as the clearest indicator of this coupling.

Figure 3(a) exhibits the temperature dependent magnetization under ZFCW, FCC and FCW protocols. Under $H_{DC}$ = 100 Oe, the three protocols exhibit the long-range magnetic ordering suggesting the canted anti-ferromagnetism with Néel temperature at $T_N \sim 173.4$ K which is confirmed by the $\frac{d(\chi T)}{dT}$ vs T with Curie-Weiss linear fit that ventures into the negative temperature scale and meets the X-axis at Curie-Weiss temperature, $\Theta_D \sim -28.6$ K by extrapolating the linear fit at the high temperature paramagnetic behaviour [30]. Long-range magnetic ordering lead by the Cr$^{3+}$ sublattice configured in the G-type CAFM structure identified with the Bertaut's notation $\Gamma_4\ (G_x, A_y, F_z)$ as shown in Fig. 3 [16]. The negative magnetization popped up during the FCC condition meets the compensation point at $T_{Comp} \sim 121$ K. Beyond the $T_{Comp}$, the Gd$^{3+}$ and Ce$^{3+}$ contribution increases which is evident through the increase in negative magnetization and attains a maximum value of $34 \times 10^{-2} \frac{\mu_B}{f.u.}$ under $H_{DC}$ = 100 Oe. From Fig. 3(b), under an applied magnetic field of 1 kOe, the magnetization in negative scale suddenly switches to the positive under FCC condition which can be understood as spin-flip transition which is well-known for GdCrO$_3$ polycrystalline system [4]. Beyond the Spin-flip transition temperature, $T_{SF} \sim 10$ K the magnetic sublattice gains the Zeeman energy which is helpful for flipping the spins from $\Gamma_4\ (G_x, A_y, F_z)$ to $\Gamma_4'\ (G_x, A_y, F_z)$ from $(00\bar{1})$ to $(001)$ where, the c-axis is the easy axis [3]. The parameters and their influence on Zeeman energy of the system is given by the following Eq. (8):

$$E_{zeeman} = -\mu_0 M_{Net} H_{Ext} \cos\theta \qquad 8$$

where $\theta$ is the angle between the $M_{Net}$, net magnetization and $H_{Ext}$, externally applied magnetic field. Net moments with $\theta = \pi$ contains large Zeeman energy and they are easy to be flipped, on the other hand the spins with $\theta < \pi$ will be flipped by applying more external magnetic field. The spins flipped increase the magnetization in the positive scale by crossing the magnetization compensation. During the FCW condition, the curve cannot reverse into the $\Gamma_4\left(G_x, A_y, F_z\right)$ configuration but continue in the $\Gamma_4'\left(G_x, A_y, F_z\right)$ condition and attains PM configuration after the $T_N$. Spin-flip phase transition occurred in this system is irreversible similar to the GdCrO$_3$ and CeCrO$_3$. From Fig. 3(c), the magnetization measurements performed under FCC at $H_{DC} \sim -200$ Oe and FCW at $H_{DC} \sim +200$ Oe in order to understand the behaviour of magnetic ordering through the applied field polarity. While field cooled cooling protocol the magnetic ordering is exactly opposite to the magnetic ordering given in the Fig. 3(a), signifying the parallel alignment of the Cr$^{3+}$ sublattice to the applied field and anti-parallel alignment of the Gd$^{3+}$ and Ce$^{3+}$ moments to the local-field applied by the Cr$^{3+}$ sublattice. At the frozen state of the magnetic-moments at 3 K, the field is changed from -200 Oe to +200 Oe and the magnetization is recorded while warming till temperatures above $T_N$. An intriguing trend has been evident through positive magnetization along with an offset from the FCC magnetization shown all through the FCW protocol above the $T_N$ even into the PM region. This very behaviour shows the characteristic pinned AFM between the Cr$^{3+}$ and Gd$^{3+}$/Ce$^{3+}$ magnetic sublattices irrespective of the applied magnetic field polarity.

In order to understand the field dependent magnetic sublattice behaviour and the stability of the spins that gets flipped, time stamp magnetization measurements at 3 K at different magnetic fields, $H_{DC}$ = 100 Oe, 200 Oe, 400 Oe, 500 Oe, 700 Oe, 1 kOe, 1.3 kOe, and 2 kOe under FCC condition as shown in the Fig. 4(a). The overall magnetization M(T) undergoing the second transition at the low temperatures associated with spin-flip transition triggered by the critical field, $H_C$ = 200 Oe but not enough to fully establish a spin-flipped magnetic sublattice. The gradual spin-flip transition is tailored by gradually changing the external magnetic fields and witness a completely compensating sublattice's spin-flip configuration from 1 kOe and above magnetic fields. The number of spins flipped depends on the amount of magnetic field applied which is clearly seen from Eq. 7. The sample has been cooled under an applied filed of 100 Oe to 3 K where the negative magnetization persists as the Zeeman energy is not enough to flip the sublattice into $\Gamma_4'$ configuration. Magnetization has been measured for 300 secs followed by measuring the magnetization at 200 Oe for 300 secs and increasing the field to 400 Oe, 700 Oe, 1 kOe and 2 kOe. The number of spins flipped and magnetization shooting up to the positive scale explicitly observed through the time stamps confirming their stability and shows the robust dependency of the Zeeman energy on the externally applied magnetic field. The path-independent nature of the spin flip transition has been observed through the cooling and heating under different magnetic fields, CHUF protocol as shown in the Fig. 4(b) [31]. The magnetization recorded under the FCC condition by applying a magnetic field of $H_{DC}$ = 200 Oe. This reflects the absence of the spin-flip transition. At 3 K the magnetic field has been raised to 70 kOe, and the FCW curve has been recorded while warming. The magnetization curve during the warming cycle reflects the completely flipped spins containing the $\Gamma_4'$ configuration.

The stability and reversibility of magnetization polarity in GCCO were examined through time-dependent magnetization measurements. Figure 4(c) shows the temperature-controlled switching under an applied field of 100 Oe in the FCC protocol. The sample was first cooled to 120 K, below $T_{Comp}$, where negative magnetization appears, and the magnetization was recorded for 300 s. The temperature was then raised to 140 K (above $T_{Comp}$) and measured for another 300 s. This cycle following 120 K → 140 K → 120 K → 140 K was repeated multiple times. Across each temperature transition, the magnetization consistently switched between negative and positive values without decay, maintaining stability for more than 4000 s. Figure 4(d) presents field-controlled switching at 120 K under FCC conditions. The sample was held at 100 Oe for 300 s, followed by an increase to 400 Oe for 300 s, and then returned to 100 Oe all at constant temperature. This sequence was repeated five times. The magnetization reversibly follows the applied field, demonstrating stable spin alignment and switching behaviour. Since the system remains above $T_{SF}$, it avoids entering the spin-flipped $\Gamma_4'$ configuration, enabling consistent switching. These results confirm reliable, repeatable magnetization polarity control, highlighting GCCO's potential for magnetic switching device applications [27, 32, 33].

In order to investigate the influence of the *A*-site doping at on the magnetocaloric effect (MCE), we executed the first quadrant ZFCW isotherm hysteresis curves at different temperatures and within the field range from 0 Oe to +90 kOe. The experimental data is plotted in Fig. 4(e) for specific temperatures within 3 K to 20 K which includes the $T_{SR}$ region. The MCE shows a direct proportional relation with the temperature derivative of magnetization ($\partial M/\partial T$) [34]:

$$\Delta S_M = \sum_i \frac{M_{i+1\,(T_{i+1},H)} - M_{i\,(T_i,H)}}{T_{i+1} - T_i} \Delta H \qquad 9$$

$$\Delta S_M = \int_0^H \left(\frac{\partial M(T,H)}{\partial T}\right) dH \qquad 10$$

with $\Delta S_M$ as the isothermal magnetic entropy change. From the above equation, it can be easily interpreted that higher the value of ($\partial M/\partial T$), larger the MCE value.

On integrating the ($\partial M/\partial T$) with respect to chosen field values, the $\Delta S_M$ curves are plotted in Fig. 4(f). We observed the single polarity dominated curves with a hint of phase transition near the temperature 3 K starting at higher field value of 30 kOe which perfectly matches with the temperature dependent magnetization data. We can see the increment in $\Delta S_M$ value with the increase in magnetic field $\Delta H$ with the broadening of the peaks near the $T_{SR}$ asymmetrically and shifts towards higher temperature with the increasing $\Delta H$. We tabulated the comparative $\Delta S_M$ values with the similar systems previously reported in Table 2. Gd has already attained great attention for its higher MCE value as well as the potential candidate for the magnetocaloric refrigeration with the $\Delta S_M$ value of 10.2 J/kg-K [35]. The pristine GdCrO$_3$ polycrystalline material holds $\Delta S_M$ value as large as 31.6 J/kg-K under $\Delta H=$ 70 kOe at 5 K [36] and hence the highest till date among the Gd systems. In our sample we attained a value of 40.70 J/kg-K under $\Delta H$ = 70 kOe at 3 K and even higher (45.27 J/kg-K) at higher $\Delta H$ = 90 kOe which is the new maximum among the rare earth orthochromite and even among the Gd system in other crystal systems as well. Here, we can infer those numerous factors can control the MCE value for this A-site doped distorted system. One of these factors can be: this sample shows higher magnetization than the similar compounds with small coercive field leading to preserving more energy during the thermal process. Other than this, other elements possess higher magnetic moment (Dy ~ 10.63 $\mu_B$) [12] but as the $\Delta S_M$ depends on the thermal derivative, the slope for this particular sample is higher resulting to enhancement in $\Delta S_M$. The magnetic spin reorientation near $T_{SR}$ (~ 3K) corresponds to the ordering of the Gd$^{3+}$ spins due to the Gd$^{3+}$-Gd$^{3+}$ interaction greatly influences the $\Delta S_M$ value.

## 4. Conclusions

In this work, we demonstrate how the typical magnetic landscape of GdCrO$_3$ can be manipulated in field, temperature and time domains. Under $H_{DC}$, the Cr$^{3+}$ sublattice creates a local field to which the $R^{3+}$ sublattice aligns anti-parallel and compensates the magnetization of the Cr$^{3+}$ sublattice which in turn leads to negative magnetization ($M_N \sim$ -7.15 emu/g at $T$ = 3 K). The overall magnetization undergoes a second anomalous change across the low temperatures associated with spin-flip transition activated by the critical field, $H_C$ = 200 Oe at $T_{SF}$ = 10 K obtained from the Zeeman interaction term. The distortions brought to reduce the Cr-O-Cr bond angle in the crystal structure by introducing a heavier cation Ce$^{3+}$ at the $A$-site which give rise to an intriguing magneto-caloric effect with change in magnetic entropy as large as 42 J/Kg-K (-$\Delta S_M$). Fine tuning of the magnetic landscapes represented by the $\Gamma_4(G_x, A_y, F_z; F_z^R)$ and $\Gamma_4'$ w.r.t. magnetic field and temperature showcase the capability of these candidates in thermo-magnetic switches, magnetic refrigeration, and other spintronic devices.

## 5. Acknowledgments

R.K.D. thank the Council of Scientific and Industrial Research, Ministry of Science and Technology, the Ministry of Education, Government of India.

7. Funding

R. K. D. and S.D. acknowledge the FIST program of the Department of Science and Technology, India, for partial support of this work (Grant Nos. SR/FST/PSII-020/2009 and SR/FST/PSII-037/2016). S. T. acknowledges the DST SERB Core Research Grant File No. CRG/2022/006155 for the support of this work. S. T. acknowledges the Central Instrument Facility (CIF) of the Indian Institute of Technology Guwahati for partial support of this work. S. T. acknowledges the invités professor fellowship from the Université of Caen. S.T. thanks the North East Centre for Biological Sciences and Healthcare Engineering (NECBH) of the Indian Institute of Technology Guwahati, Assam, for partial support of this work under the grant DBT Project code: BT/NER/143/SP44675/2023.

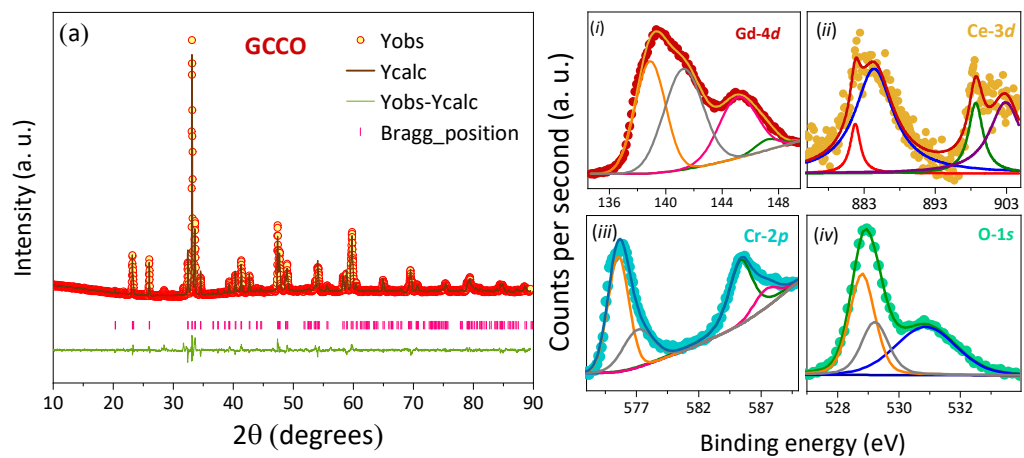

**Figure 1.** (a) Rietveld refined X-ray diffraction patterns of GCCO showing single phase *Pbnm* orthorhombic perovskite structure. X-ray photoelectron spectroscopy of GCCO (*i*) Gd − 4$d$, (*ii*) Ce − 4$d$, (*iii*) Cr − 2$p$ (*iv*) O − 1$s$. Scattered symbols represent the original data and solid lines are the fitted curves.

**Table 1** The refined crystallographic parameters obtained from XRD of GCCO. The parameters, *a, b, c, and V* are lattice constants and lattice volume respectively. $\theta_1$ and $\theta_2$ are the Cr-O$_{(1)}$-Cr and Cr-O$_{(2)}$-Cr bond angles. $\theta$ and $\phi$ are the in-phase and out-of-phase tilt angles *w.r.t.* 110] and [001] respectively. $\Delta$, octahedral distortion, *t*, tolerance factor, $r_{avg}$, average radius of the $R^{3+}$ cation.

| Parameter | | Atomic positions | | Bond lengths | | Bond angles | | Other parameters | |
|---|---|---|---|---|---|---|---|---|---|
| *a* | 5.31946 | Gd/Ce | 0.99365 | Gd/Ce-O$_{(1)}$ | 2.3256(0) | $\theta_1$ | 155.3 | $\theta$ [110] | 12.36 |
| *b* | 5.50727 | (4c) | 0.05841 | (*two*) | 2.3779(0) | | | $\phi$ [001] | 14.92 |
| *c* | 7.60545 | | 0.25000 | | | $\theta_2$ | 145.5 | $\theta$ | 8.45 |
| *V* | 222.807 | | 1.000 | Gd/Ce-O$_{(2)}$ | 2.6429(0) | | | | |
| | | Cr | 0.00000 | (*six*) | (*two*) | | | (*from lattice parameters*) | |
| | | (4b) | 0.50000 | | 2.5845(0) | | | | |
| | | | 0.00000 | | (*two*) | | | $\phi$ | 15.01 |
| | | | 1.000 | | 2.2850(0) | | | | |
| | | O1 | 0.07624 | | (*two*) | | | (*from lattice parameters*) | |
| | | (4c) | 0.48276 | Cr-O$_{(1)}$ | | | | $r_{avg}$ | 1.062 |
| | | | 0.25000 | (*two*) | 1.9464(0) | | | | |
| | | | 0.900 | | (*two*) | | | *t* | 0.8663 |
| | | O2 | 0.68067 | Cr-O$_{(2)}$ | | | | | |
| | | (8b) | 0.29345 | (*four*) | 1.9239(0) | | | $\Delta$ ($\times 10^{-4}$) | 1.2800 |
| | | | 0.05355 | | (*two*) | | | | |
| | | | 1.599 | | 2.0845(0) | | | s | 0.03469 |
| | | | | | (*two*) | | | $\phi$ | 17.1748 |

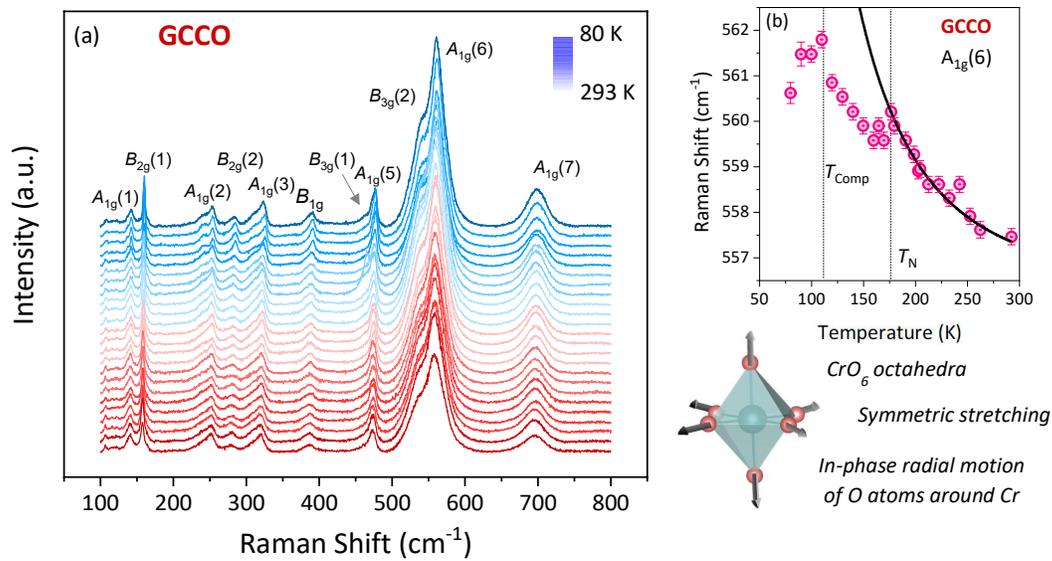

**Figure 2** (a) Raman spectra with all the existing modes indexed for GCCO from 80 K to 293 K. Self-explanatory schematic of symmetric stretching mode, along with temperature dependence of $A_{1g}(6)$ mode.

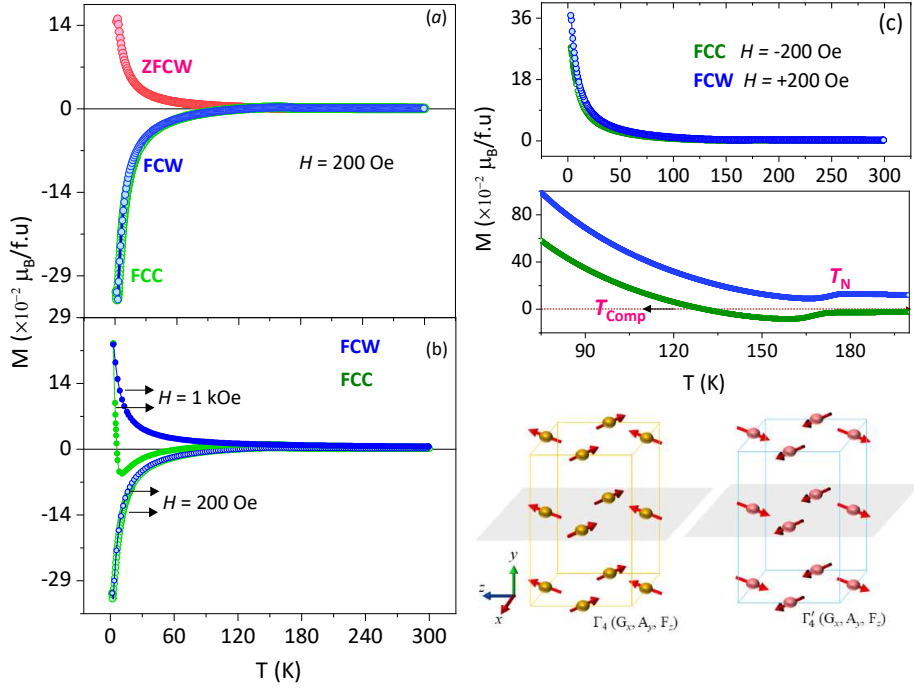

**Figure 3** (*a*) Temperature dependent magnetization of GCCO under (a) ZFCW, FCC, and FCW protocols. (*b*) FCC and FCW at field 1kOe and 200 Oe. Irreversible Spin-flip transition observed during the FCC at 1 kOe. (*c*) FCC under the applied field $H = -200$ Oe and FCW under the field $H = +200$ Oe. Zoomed view given below to show the $T_{Comp}$ observed only under the FCW condition.

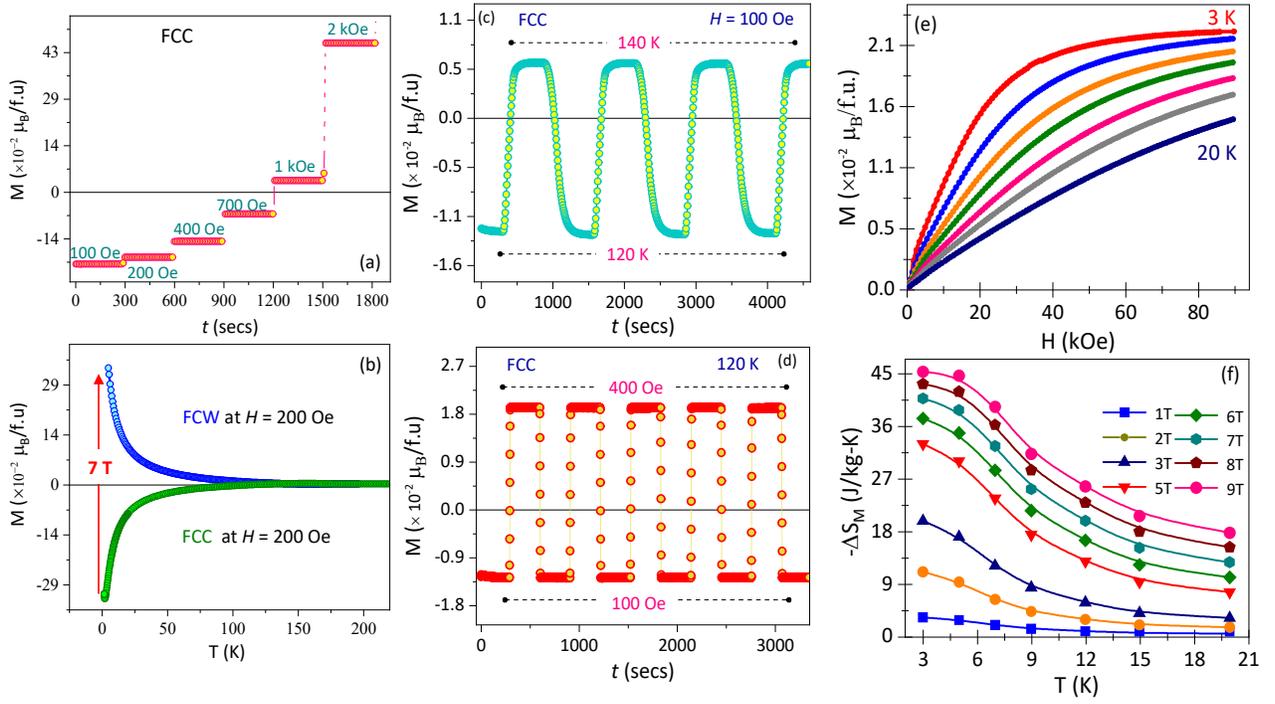

**Figure 4** (*a*) Magnetization versus time measurements at different magnetic fields 100 Oe, 200 Oe, 400 Oe, 500 Oe, 700 Oe, 1 kOe, 1.3 kOe, 2 kOe representing the gradual increase in the no. of spins that flip *w.r.t.* applied field. (*b*) Temperature dependent magnetization under the FCC condition by applying external field of $H$ = 200 Oe. Raise the magnetic field to 70 kOe at 5 K. Followed by decreasing the magnetic field again to 200 Oe and measurement was recorded under FCW condition. Time stamp measurements of (*c*) temperature dependent magnetization switching of GCCO between 120 K and 140 K under FCC condition at applied field, $H$ = 100 Oe, (*d*) field dependent magnetization switching between 400 Oe and 100 Oe under FCC condition at 120 K. (*e*) Magnetic field dependence of magnetization of GCCO at different temperatures from 3 K to 20 K under ZFC. (*f*) Temperature dependence of isothermal entropy change, $-\Delta S_M$.

**Table 2** $-\Delta S_M$, RCP of various potential magnetic refrigerant materials having operating temperatures below 20 K along with $GdCrO_3$.

| Material | $\mu_0H$ (T) | $-\Delta S_M$ (J/kg-K) | T (K) | RCP (J/kg) | Reference |
|---|---|---|---|---|---|
| Gd | 5 | 10.2 | - | 410 | [37] |
| $Gd_{0.9}Ce_{0.1}CrO_3$ | 5 | 32.9 | 3 | | This work |
| $GdCrO_3$ (single crystal) | 4 | 26.1 | 5 | - | [38] |
| $Gd_{0.9}Ce_{0.1}CrO_3$ | 6 | 37.3 | 3 | | This work |
| $La_{0.7}Ca_{0.3}MnO_3$ | 2 | 2.2 | | 55 | [39] |
| $La_{0.67}Sr_{0.1}Ca_{0.23}MnO_3$ | 5 | 6 | - | 278.55 | [40] |
| $HoCrO_3$ | 7.2 | 7 | 20 | 408 | [41] |
| $DyCrO_3$ | 4 | 8.4 | 15 | 217 | [42] |
| $GdCrO_3$ (polycrystalline) | 7 | 36.9 | 5 | 542 | [4] |
| $Gd_{0.9}Ce_{0.1}CrO_3$ | 7 | 40.7 | 3 | | This work |
| $GdMnO_3$ | 8 | 31 | 19 | - | [43] |
| $Gd_{0.9}Ce_{0.1}CrO_3$ | 9 | 45.3 | 3 | | This work |
| $HoMnO_3$ | 7 | 12.5 | 10 | 312 | [28] |
| $ErCrO_3$ | 7 | 10.7 | 15 | 416 | [35] |
| $Er_{0.33}Gd_{0.67}CrO_3$ | 7 | 27.6 | 5 | 252 | [35] |
| $TbCrO_3$ | 4.5 | 12.2 | 4.5 | 125 | [20] |
| $DyCr_{0.7}Fe_{0.3}O_3$ | 7 | 13.1 | 5 | 500 | [44] |
| $HoFeO_3$ (single crystal) | 7 | 19.2 | 4.5 | 220 | [45] |
| $HoCr_{0.7}Fe_{0.3}O_3$ | 7 | 6.8 | 20 | 387 | [44] |
| $Ho_{0.67}Tm_{0.33}CrO_3$ | 7 | 6.7 | 17 | - | [41] |
| $La_{0.6}Ca_{0.4}MnO_3$ | 5 | 8.3 | 61.2 | 508 | [46] |
| $EuTi_{0.9}Cr_{0.1}O_3$ | 2 | 30 | 4.2 | 125 | [47] |
| $La_{0.7}(Sr, Ba)_{0.3}MnO_3$ | 5 | 2.8 | 103.8 | 285.8 | [48] |
| $La_{0.57}Nd_{0.1}Sr_{0.18}Ag_{0.15}MnO_3$ | 5 | 5.1 | | 146.74 | [49] |